\documentclass[twocolumn,journal, onecolumn]{IEEEtran}
\usepackage[T1]{fontenc}
\usepackage[latin9]{inputenc}
\usepackage{color}
\usepackage{units}
\usepackage{amstext}
\usepackage{graphicx}
\PassOptionsToPackage{normalem}{ulem}
\usepackage{ulem}

\makeatletter
\usepackage[caption=false,font=footnotesize]{subfig}

\makeatother

\begin{document}

\title{A New Wireless Communication Paradigm through Software-controlled
Metasurfaces }

\author{Christos Liaskos\IEEEauthorrefmark{1}, Shuai Nie\IEEEauthorrefmark{2},
Ageliki Tsioliaridou\IEEEauthorrefmark{1}, Andreas Pitsillides\IEEEauthorrefmark{3},
Sotiris Ioannidis\IEEEauthorrefmark{1}, and Ian Akyildiz\IEEEauthorrefmark{2}\IEEEauthorrefmark{3}\\
 {\small{}\IEEEauthorrefmark{1}Foundation for Research and Technology
- Hellas (FORTH)}\\
 {\small{}Emails: \{cliaskos,atsiolia,sotiris\}@ics.forth.gr}\\
 {\small{}\IEEEauthorrefmark{2}Georgia Institute of Technology, School
of Electrical and Computer Engineering}\\
 {\small{}Emails: \{shuainie, ian\}@ece.gatech.edu}\\
 {\small{}\IEEEauthorrefmark{3}University of Cyprus, Computer Science
Department}\\
 {\small{}Email: Andreas.Pitsillides@ucy.ac.cy}}
\maketitle
\begin{abstract}
Electromagnetic waves undergo multiple uncontrollable alterations
as they propagate within a wireless environment. Free space path loss,
signal absorption, as well as reflections, refractions and diffractions
caused by physical objects within the environment highly affect the
performance of wireless communications. Currently, such effects are
intractable to account for and are treated as probabilistic factors.
The paper proposes a radically different approach, enabling deterministic,
programmable control over the behavior of the wireless environments.
The key-enabler is the so-called HyperSurface tile, a novel class
of planar meta-materials which can interact with impinging electromagnetic
waves in a controlled manner. The HyperSurface tiles can effectively
re-engineer electromagnetic waves, including steering towards any
desired direction, full absorption, polarization manipulation and
more. Multiple tiles are employed to coat objects such as walls, furniture,
overall, any objects in the indoor and outdoor environments. An external
software service calculates and deploys the optimal interaction types
per tile, to best fit the needs of communicating devices. Evaluation
via simulations highlights the potential of the new concept.
\end{abstract}


\section{Introduction\label{sec:Intro}}

\IEEEPARstart{W}{ireless} communications are rapidly evolving towards
a software-based functionality paradigm, where every part of the device
hardware can adapt to the changes in the environment. Beamforming-enabled
antennas, cognitive spectrum usage, adaptive modulation and encoding
are but a few of the device aspects that can now be tuned to optimize
the communication efficiency~\cite{Akyildiz.2016}. In this optimization
process, however, the environment remains an uncontrollable factor:
it remains unaware of the communication process undergoing within
it. In this paper we make the environmental effects controllable and
optimizable via software.

A wireless environment is defined as the set of physical objects that
significantly alter the propagation of electromagnetic (EM) waves
among communicating devices. In general, emitted waves undergo attenuation
and scattering before reaching an intended destination. Attenuation
is owed to material absorption losses, and the natural spreading of
power within space, i.e., the distribution of power over an ever-increasing
spherical surface. Wave scattering is owed to the diffraction, reflection
and refraction phenomena, which result into a multiplicity of propagation
paths between devices. The geometry, positioning and composition of
objects define the propagation outcome, which is, however, intractable
to calculate except for simple cases.
\begin{figure*}[t]
\begin{centering}
\includegraphics[width=1\textwidth]{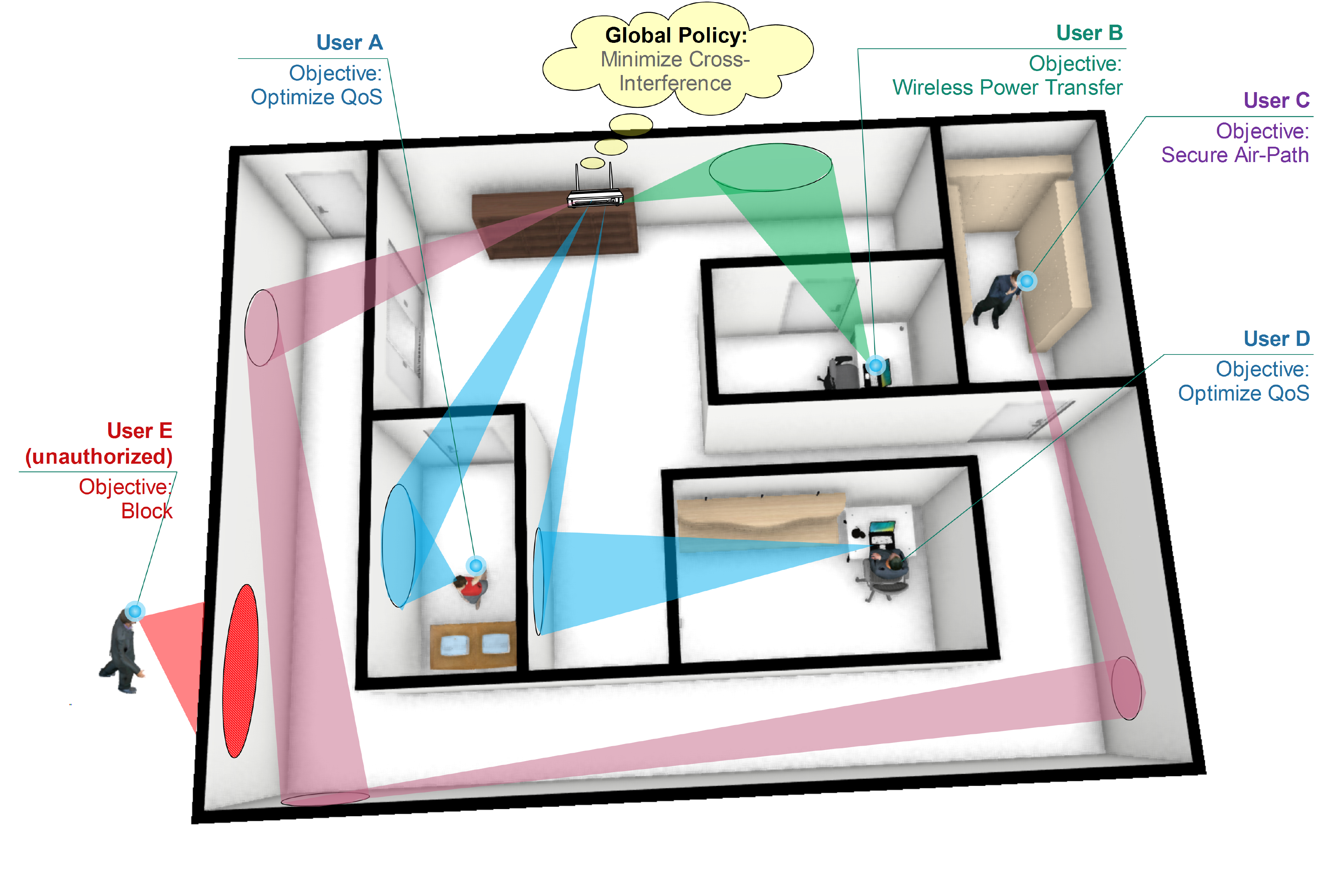}
\par\end{centering}
\caption{\label{fig:WirCommExample}Programmable wireless environments can
exhibit user-adapting, unnatural wireless behavior, manipulating EM
waves to match the requirements of users. Wireless power transfer,
Quality of Service (QoS) and Security scenarios are exemplary illustrated.}
\end{figure*}

Apart from being uncontrollable, the environment has a generally negative
effect on the communication efficiency. The signal attenuation limits
the connectivity radius of nodes, while multi-path propagation results
into fading phenomena, a well-studied effect which introduces drastic
fluctuations in the received signal power. The signal deterioration
is perhaps the major consideration in forthcoming mm-wave and THz
communications. While these extremely high communication frequencies
offer unprecedented data rates and device size minimization, they
suffer from acute attenuation owed to molecular absorption, multi-path
fading and Doppler shift even at pedestrian speeds, limiting their
present use in short line-of-sight distances~\cite{Akyildiz.2016}.
Existing mitigation approaches propose massive MIMO and 3D beamforming
at the device-side~\cite{Kim.2014b}, and passive reflectors/active
reflectarrays carefully placed at intermediate points within a space~\cite{reflectInfocom.2017,PADOSarxiv}.
However, while these approaches provide a good degree of control over
the directivity of wireless transmissions, they pose mobility and
hardware scalability issues. Moreover, the control is limited to directivity
and does not extend to full EM manipulation. As a result, the wireless
environment as a whole remains unaware of the ongoing communications
within it, and the channel model continues to be treated as a probabilistic
process, rather than as a \emph{software-defined service}.

The key-enabler for building a programmable wireless environment is
the concept of metamaterials and metasurfaces~\cite{Yang.2016}.
Metamaterials are artificial structures, with engineered EM properties
across any frequency domain. In their most common form, they comprise
a basic, simple structure, the \emph{meta-atom}, which is repeated
periodically within a volume. Metasurfaces are the 2D counterparts
of metamaterials, in the sense of having small-but not negligible-depth.
While materials found in the nature derive their properties from their
molecular structure, the properties of metamaterials stem from the
form of their meta-atom design. Thus, when treated macroscopically,
metamaterials exhibit \emph{custom permittivity and permeability}
values locally, even beyond those found in natural materials. As a
consequence, metamaterials enable exotic interactions with impinging
EM waves, being able to fully \emph{re-engineer} incoming waves. Finally,
the dynamic meta-atom designs can be altered with simple external
bias\textendash such as a binary switch\textendash endowing metamaterials
and metasurfaces with adaptivity. The naming of metamaterials is a
testament to their simple and scalable internal structure, which classifies
them as materials rather than as antenna arrays.

The methodology for introducing software control over the EM behavior
of a wireless environment consists at coating objects, such as walls,
furniture, overall any objects in the indoor or outdoor environments,
with HyperSurfaces, a forthcoming class of software-controlled metasurfaces~\cite{TheVISORSURFproject.2017}.
HyperSurfaces merge networked control elements with adaptive metasurfaces.
The control elements apply the proper bias to adaptive metasurface
meta-atoms, thereby attaining a desired macroscopic EM behavior. Additionally,
the HyperSurface has interconnectivity capabilities, which allow it
to enter control loops for adapting their performance. In this paper
we introduce the HyperSurface tile architecture and the process of
using them to build programmable wireless environments. We discuss
the high-level programming interfaces for interacting with tiles,
and detail the enabling of a new class of software that will treat
wireless propagation as an application. We proceed further to study
the practical incorporation to existing networking infrastructures
and to evaluate the novel capabilities of the programmable environments
via raytracing-based simulations.

The remainder of this paper is organized as follows. Section~\ref{sec:App}
presents the programmable wireless environment concept and Section~\ref{sec:arch}
details its architecture. Evaluation via simulations takes place in
Section~\ref{sec:Evaluation}. Research challenges are discussed
in Section~\ref{sec:future}, and the paper is concluded in Section~\ref{sec:Conclusion}.

\section{Programmable wireless environments: the concept \label{sec:App}}

Consider a scenario of wireless communications within a space, as
shown in Fig.~\ref{fig:WirCommExample}. Several users require connectivity,
each with different requirements. Users A and D are interested in
optimal connection quality, user B is interested in wireless power
transfer, and user C requires eavesdropping avoidance measures. Finally,
user E represents unauthorized access or interference attempts, which
may be deliberate or random. In the common, passive environment, such
objectives cannot be met efficiently. Devices employ beamforming to
find promising wave transmission directions, but the environment remains
oblivious to the process. EM waves scatter uncontrollably upon objects,
sharply losing their focus and carried power, causing interference,
performance drop and security concerns.

In the case of a programmable wireless environment, objects such as
walls ceilings, etc., receive HyperSurface-tile coating that enables
them to re-engineer impinging waves in a software-defined manner.
Each tile incorporates a lightweight Internet-of-Things (IoT) gateway
which enables it to receive commands from a central configuration
service and set its custom EM behavior accordingly. In collaboration
with existing device beamforming mechanisms and location discovery
services, the programmable environment allows for novel capabilities,
essentially treating EM propagation in a manner reminiscent of routers
and firewalls in classical networking. As shown in Fig.~\ref{fig:WirCommExample},
users A and D receive maximum signal-to-interference power levels
by carefully focusing the EM waves in a lens-like manner and avoiding
mutual interference. Moreover, the wave propagation is groomed further
to achieve constructive superposition at the user devices, optimizing
their power-delay profile (PDP) and avoiding the negative effects
of multi-path fading. The environmental response for user B targets
maximum wireless power transfer using a combination of custom wave
steering and focusing, but without PDP concerns. For user C, the environment
establishes a ``private air-route'', that avoids all other users
to reduce the risk of eavesdropping. Finally, the unauthorized user
E is blocked by instructing the environment to absorb his emissions,
potentially using the harvested energy in a constructive way, such
as powering some HyperSurface tiles.

We proceed to detail the architecture of the HyperSurface tiles that
comprise a programmable environment. Moreover, we discuss its incorporation
to existing network infrastructures, as well as the process for modeling
and treating \emph{wireless propagation as an app}.

\section{The Architecture of Programmable Environments\label{sec:arch}}

We begin by presenting some prerequisite knowledge on the structure
and properties of metasurfaces. Here we focus on the basics required
to subsequently describe the HyperSurfaces.

\begin{figure}[t]
\begin{centering}
\includegraphics[width=1\columnwidth]{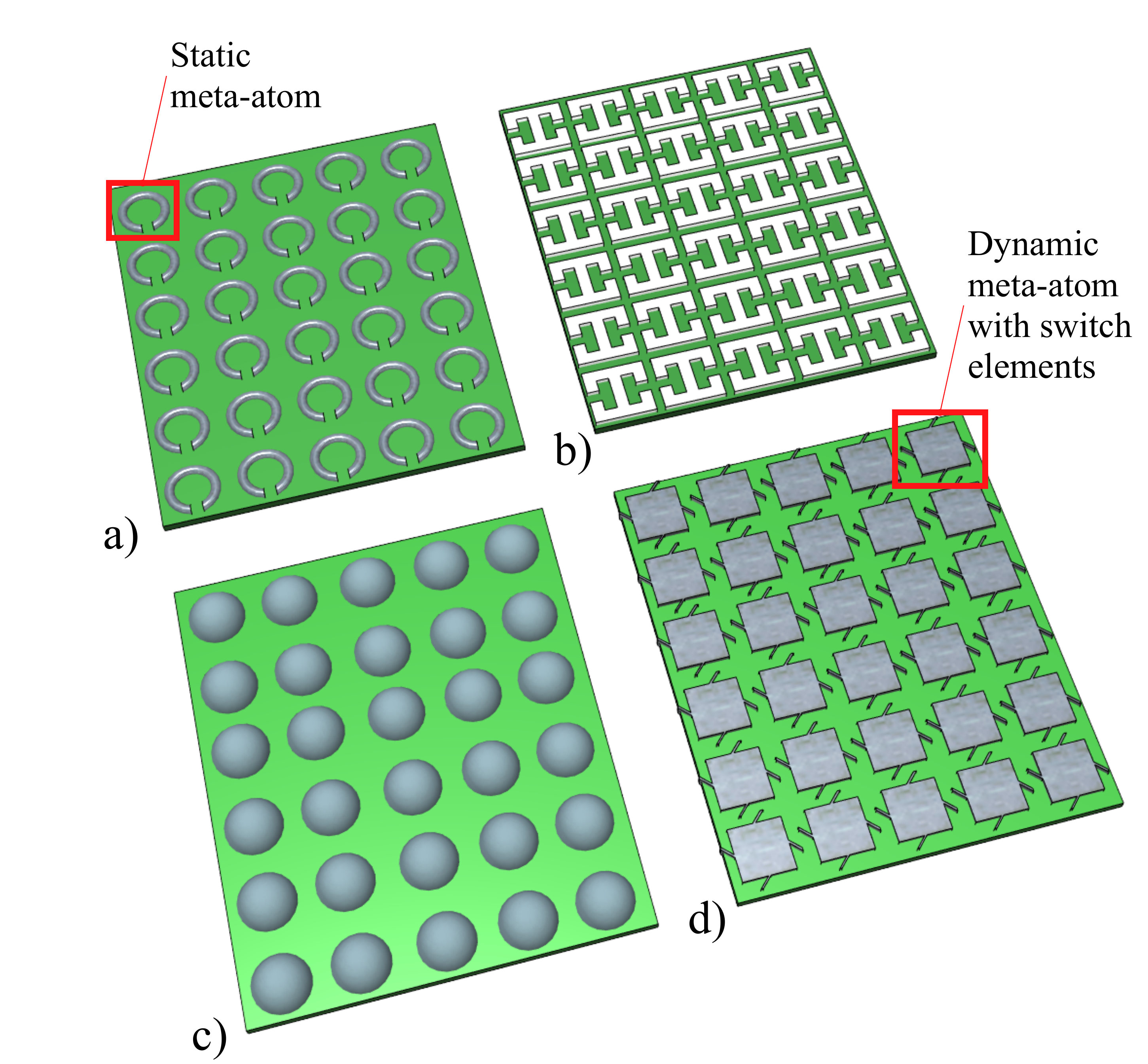}
\par\end{centering}
\caption{\label{fig:Mspatterns}Meta-atom patterns that have been commonly
employed and investigated in metasurface research. }
\end{figure}

A metasurface is a composite material layer, designed and optimized
to function as a tool to control and transform EM waves~\cite{Yang.2016}.
They commonly comprise a conductive pattern repeated over a dielectric
substrate. Examples of meta-atom patterns constituting the building
blocks of some of the most common metasurfaces are shown in Fig.~\ref{fig:Mspatterns}.
The operating principle of metasurfaces is as follows. When EM waves
impinge on a metasurface, it creates currents in it via induction.
In the case of static meta-atoms (Fig.~\ref{fig:Mspatterns}-a),
the total current pattern within the surface is fully defined by the
meta-atom geometry and composition. In dynamic designs (Fig.~\ref{fig:Mspatterns}-d),
the current pattern also depends on the states of the switching elements.
The inducted current also creates a response field, following the
laws of electromagnetism. The meta-atoms are engineered to yield a
custom response field.

The meta-atom size and the thickness of the tile are important design
factors, which define the maximum frequency for EM wave interaction.
As a rule of thumb, meta-atoms are bounded within a square region
of $\nicefrac{\lambda}{10}\leftrightarrow\nicefrac{\lambda}{5}$,
$\lambda$ being the EM interaction wavelength. The minimal HyperSurface
thickness is also in the region of $\nicefrac{\lambda}{10}\leftrightarrow\nicefrac{\lambda}{5}$.
Thus, for an interaction frequency of $5\,GHz$, the meta-atom would
have a side of $\sim8\,mm$, with similar thickness.

We note that dynamic meta-atom designs constitute an extensively studied
subject in the literature, offering a wide variety of choices. An
extremely wide array of EM interaction types (denoted as \emph{functions})
have been achieved across any spectrum, e.g. wave steering, polarizing,
absorbing, filtering and collimation resulting from fascinating metasurface
properties such as near zero permittivity and/or permeability response,
peculiar anisotropic response leading, e.g., to hyperbolic dispersion
relation, giant chirality, non-linear response and more~\cite{Yang.2016,Su.2017b,Lee.2012}.

\subsection{The HyperSurface }

A HyperSurface tile is envisioned as a planar, rectangular structure
that can host metasurface functions over its surface, with programmatic
control. It comprises a stack of virtual and physical components,
shown in Fig.~\ref{fig:HSFTileArch}, which are detailed below.

\uline{The Functionality \& Configuration layer}s. A HyperSurface
tile supports software descriptions of metasurface EM functions, allowing
a programmer to customize, deploy or retract them on-demand via a
programming interface with appropriate callbacks. These callbacks
have the following general form:
\begin{figure*}[t]
\begin{centering}
\includegraphics[width=1\textwidth]{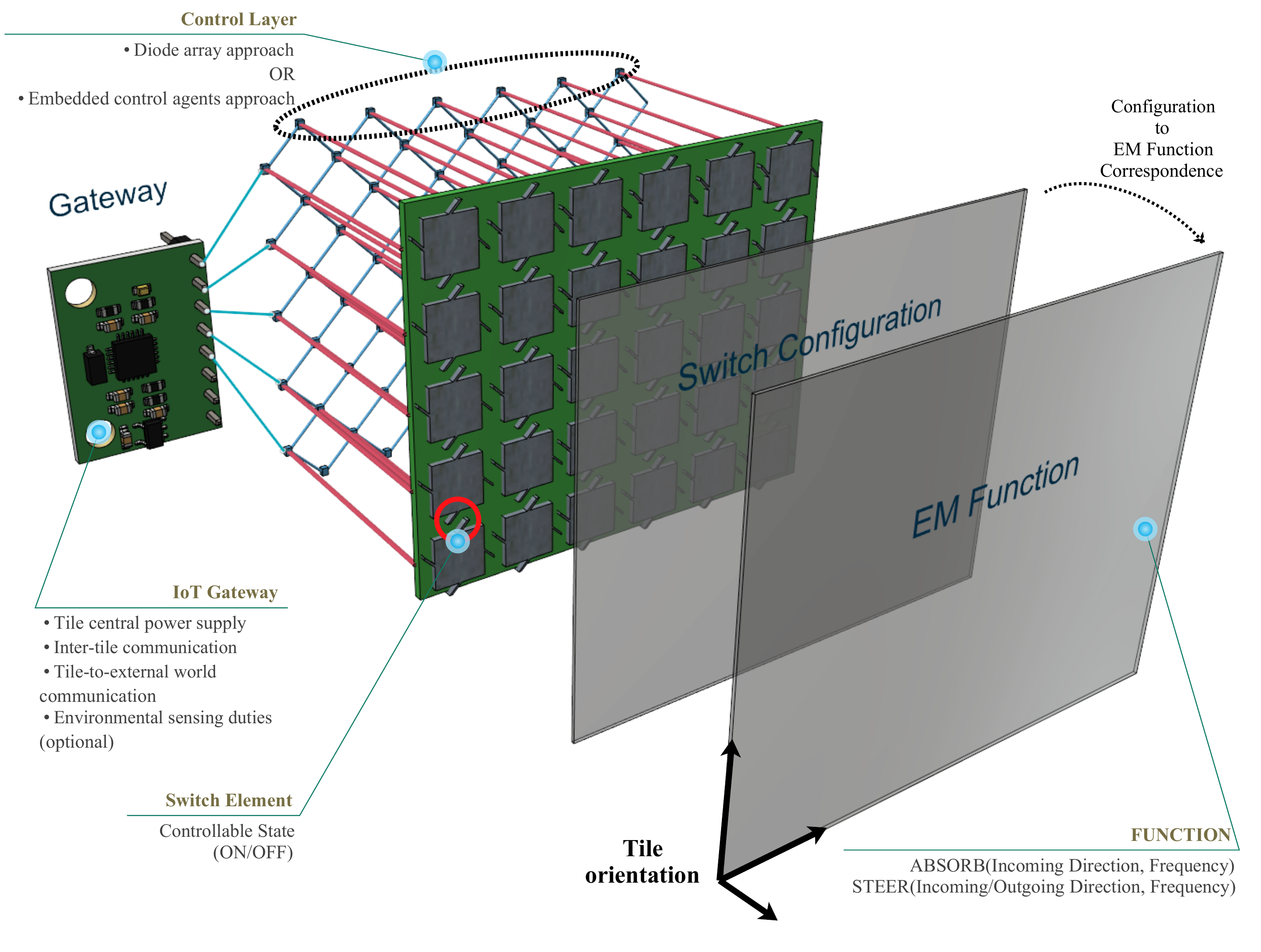}
\par\end{centering}
\caption{\textcolor{black}{\label{fig:HSFTileArch}The functional and physical
architecture of a single HyperSurface tile. A desired and supported,
EM function is attained by a switch state configuration setup. Inter-tile
and external communication, is handled by standard gateway hardware.}}
\end{figure*}

\[
\texttt{outcome}\gets\texttt{callback(action\_type, parameters)}
\]
The $\texttt{action\_type}$ is an identifier denoting the intended
function to be applied to the impinging waves, such as $\texttt{STEER}$
or $\texttt{ABSORB}$. Each action type is associated to a set of
valid parameters. For instance, $\texttt{STEER}$ commands require:
i) an incident wave direction, ii) an intended reflection direction,
and iii) the applicable wave frequency band. $\texttt{ABSORB}$ commands
require no incident or reflected wave direction parameters.

The functionality layer is exposed to programmers via an API that
serves as a strong layer of abstraction. It hides the internal complexity
of the HyperSurface and offers general purpose access to metasurface
functions, without requiring knowledge of the underlying hardware
and physics. Thus, the configuration of the phase switch materials
that matches the intended EM function is derived automatically, without
the programmer's intervention.

\uline{The Metasurface Layer}. It is the metasurface hardware comprising
dynamic meta-atoms, whose states are altered to yield and intended
EM function. This layer comprises both the passive and active elements
of meta-atoms. For instance, the example of Fig.~\ref{fig:HSFTileArch}
comprises conductive square patches (passive) and switches (active).
It is noted that even simple, ON (most conductive) / OFF (most insulating)
switches are sufficient for building metasurfaces supporting an impressive
range of EM functions~\cite{Yang.2016}.

Large area electronics (LAE) constitute very promising approaches
for manufacturing the metasurface layers~\cite{LAEBOOK}. LAE can
be manufactured using conductive ink-based printing methods on flexible
and transparent polymer films, and incorporate polymer switches (diodes)~\cite{LAEBOOK}.
Apart from minimal cost, the LAE approach favors scalability and deployment
of HyperSurfaces. Tiles can be manufactured as large films with metasurface
patterns and diode-switches printed on them, and be placed upon common
objects (e.g., glass, doors, walls, desks), which may also play the
role of the dielectric substrate for the metasurface.

\uline{The Intra-tile Control Layer.} This layer describes the
hardware components and wiring that enables the programmatic control
over the switches of the metasurface layer. A highly promising, cost-effective
and highly scalable approach, is to control the metasurface switches
as a diode array~\cite{Sekitani.2009}, i.e., as a common light-emitting-diode
(LED) display works. Meta-atoms are treated as very simple ``pixels'',
with just two ``colors'' (ON/OFF). The diode array approach results
in a very simple control layer, which comprises just the wiring to
connect each meta-atom switch to the gateway (discussed below). Moreover,
it entails a very low power drain. For instance, assume a meta-atom
with size $8\times8$~$mm$, which can interact with waves modulated
at $5$ GHz. A total number of $324,375$ meta-atoms are required
for coating a $5\times3$~$m$ wall. As shown in~\cite[p. 497]{Sekitani.2009},
a single elastic diode exhibits a drain of $5\,V\cdot1.6\,\mu A=8\,\mu W$
when powered. Thus, the total coating of the wall will drain $\sim1.88\,W$
at a \emph{maximum}\textendash i.e., when all diodes are set to 'ON'\textendash or
$125$~$\nicefrac{mW}{m^{2}}$, which constitutes a very promising
indicative value.

Apart from the presently realizable diode array control approach,
forthcoming nanonetwork technologies may also be considered as control
agents in the future~\cite{Liaskos.2015b}. Nanonetworks comprise
a network of wireless nano-sized electronic controllers, each with
responsibility over one active meta-atom element. The controllers
are able to exchange information, in order to propagate switch configuration
information within the tile. Nano-controllers are envisioned to be
autonomic in terms of power supply. While still at its early-stages,
the nanonetwork approach promotes the seamless integration of control
elements within a material, while it may also enable materials with
embedded intelligence, able to tune their EM behavior in an autonomous
fashion.

\uline{The Tile Gateway Layer.} It specifies the hardware (Gateway)
and protocols that enable the bidirectional communication between
the controller network and the external world (such as the Internet),
as well as the communication between tiles. This provides flexibility
in the HyperSurface operation workflow, as follows. In general, multiple
tiles are expected to be used as coating of large areas, as discussed
in Section~\ref{sec:App}. Moreover, the tile hardware is intended
to be inexpensive, favoring massive deployments. Based on these specifications,
existing IoT platforms can constitute promising choices for tile gateways~\cite{Akyildiz.2016}.
The sensing capabilities of existing IoT platforms may optionally
facilitate the monitoring of the tile environment, such as the impinging
wave power measurements, enabling the adaptive tuning of the tile
functions~\cite{nanocom.2017}.

The described interconnectivity approach can also be employed during
the tile design phase, for the automatic definition of the supported
EM functions and their input parameter range. Since deriving the tile
behavior via analysis is challenging in all but static meta-atom cases,
an approach based on learning heuristics can also be employed instead~\cite{Yang.2016}.
According to it, an intended EM function is treated as an objective
function and, subsequently, the tile is checked for compliance via
illumination by an external wave (input) and reflection/absorption
measurements (output). A learning heuristic is then employed to detect
iteratively the best switch configuration that optimizes the output
of the objective function. Once detected, the best configuration is
stored in a lookup table for any future use. It is noted that metasurfaces
are generally not known to exhibit an inherent limitation regarding
the EM functions that they can support.

\subsection{Incorporation to networking infrastructure}

\begin{figure}[t]
\begin{centering}
\includegraphics[width=1\columnwidth]{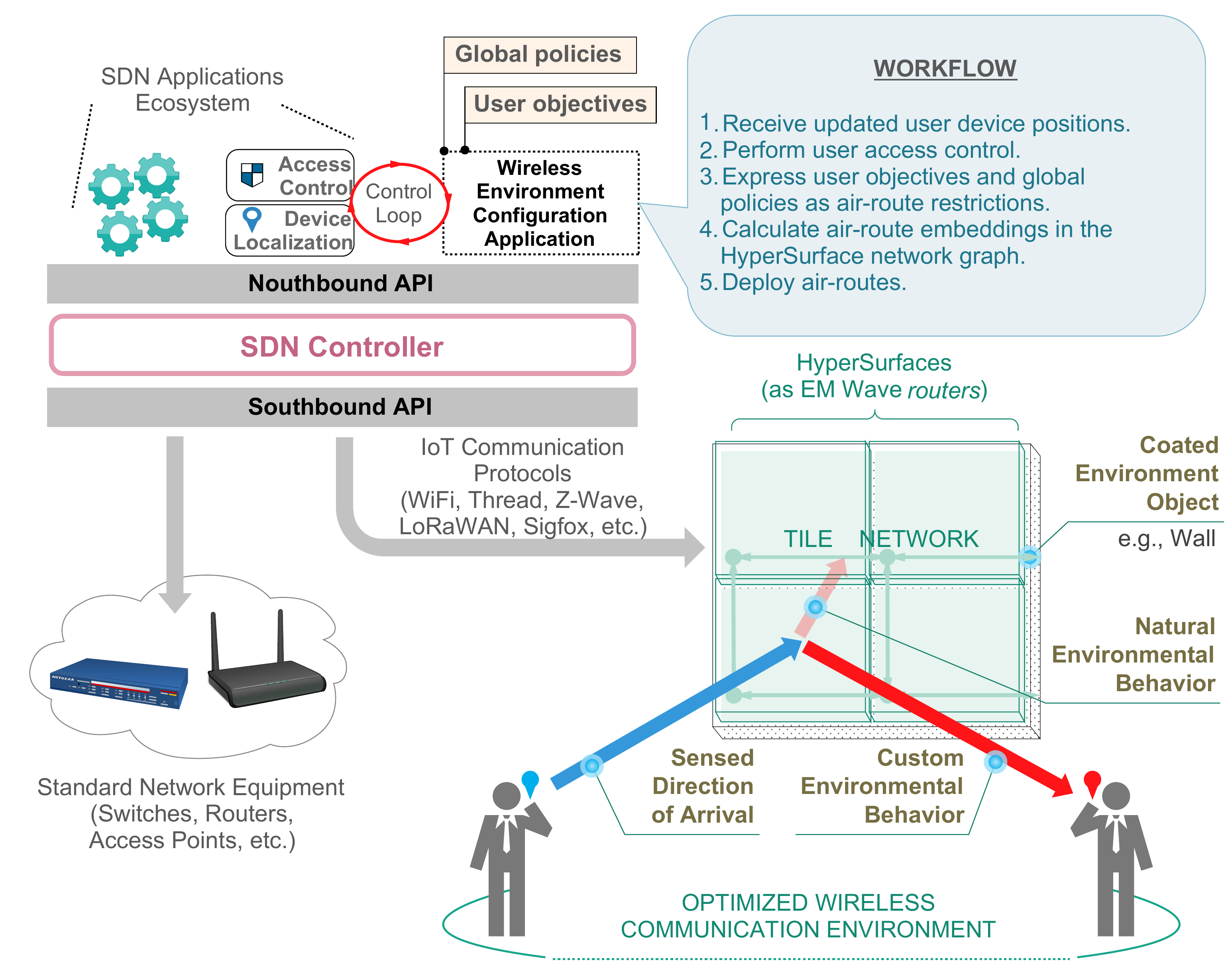}
\par\end{centering}
\caption{\label{fig:workflow}Schematic of the programmable wireless environment
incorporation principles to existing SDN infrastructures.}
\end{figure}

Programmable wireless environments can be incorporated to existing
network infrastructures without altering their workflow. Especially
in the case of Software-Defined Networks (SDN)~\cite{Akyildiz.2016},
the programmable environments can be clearly modeled as a set of software
services, as shown in Fig.~\ref{fig:workflow}. SDN has gained significant
momentum in the past years due to the clear separation it enforces
between the network control logic and the underlying hardware. An
SDN controller abstracts the hardware specifics (``southbound''
direction) and presents a uniform programming interface (``northbound'')
that allows the modeling of network functions as applications.

Using the SDN paradigm, HyperSurface tiles can be considered as wave
routing hardware. Notice that the tiles employ common IoT devices
as gateways, whose communication protocols are mainstream and typically
supported by SDN controllers. The custom environmental behavior to
serve a set of users is calculated by a wireless environment configuration
application. The application receives the device positions, the user
objectives and the global policies as inputs and calculates the fitting
air paths. A control loop is established with existing device position
discovery and access control SDN applications, constantly adapting
to changes in the security policies and user device location updates.

\subsection{Workflow of the environment configuration service}

\begin{figure}[t]
\begin{centering}
\includegraphics[width=1\columnwidth]{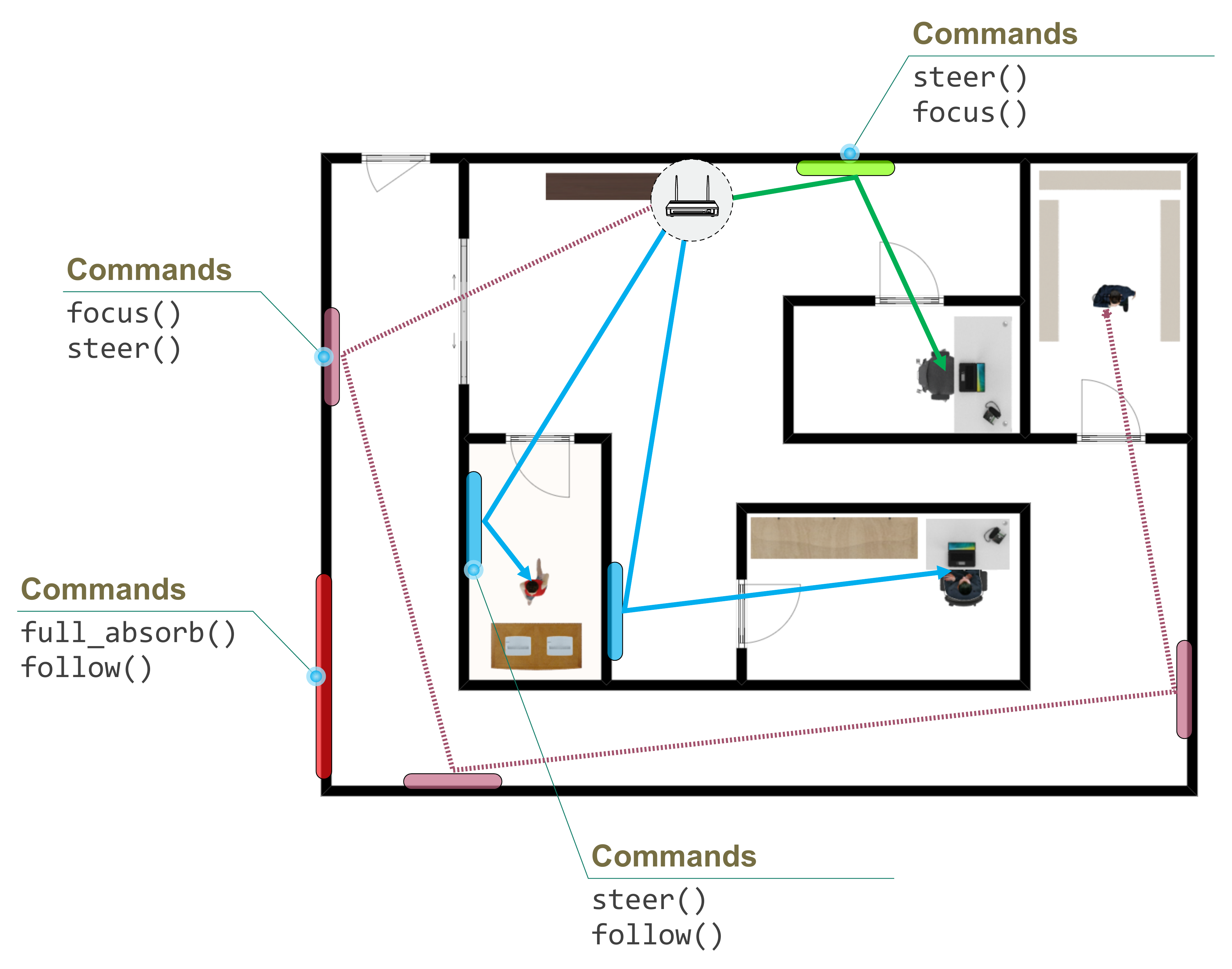}
\par\end{centering}
\caption{\label{fig:Floor2D}The wireless environment configuration process
as a routing problem. The illustrated case corresponds to the scenario
of Fig.~\ref{fig:WirCommExample}.}
\end{figure}

A HyperSurface-coated environment can treat the EM wave propagation
similar to the routing process in classic networking. Connecting two
wireless devices becomes a problem of finding a route over HyperSurface
tiles, while blocking access to a wireless device is achieved by absorbing
or deflecting its EM emissions. An example is given in Fig.~\ref{fig:Floor2D}
which studies a possible EM routing configuration to serve the objectives
of the scenario shown in Fig.~\ref{fig:WirCommExample}. Software
commands are combined and sent to the proper tile gateways, manipulating
EM waves, steering, absorbing and focusing them as needed. Finding
the air-routes that fulfill the objectives of multiple users can be
treated as a network embedding problem. When an EM beam from a device
impinges on, e.g., a wall, the affected tiles can be seen as the user
entry-points in the graph of connectable tiles. The tile graph comprises
a node for each tile and a link between any two tiles can steer EM
waves to each other. User objectives can be treated as air-route requirements,
e.g., selecting the K-Shortest paths to connect the device entry-points,
using routes that avoid other users for increased security, etc. These
requirements can then be embedded to the tile graph using well-known
techniques~\cite{EMBED}.

Having described the tile configuration process as an embedding problem,
we proceed to outline the total workflow of the configuration service.
The service forms a continuous loop with device location discovery
systems: it receives the updated locations of user devices and tunes
the behavior of the wireless environment accordingly. It is noted
that tiles can facilitate the device location discovery process~\cite{nanocom.2017}.
The existing user access mechanisms of the network infrastructure
are executed. If a device is deemed unauthorized, a ``block'' objective
is formed for it. Authorized devices that are aware of the programmable
environment can express their specific objectives by posting a request
to the configuration service. Unaware devices are treated by global
policies. The environment configuration service produces the matching
air-routes and proceed to deploy them by sending corresponding EM
manipulation commands to the tile gateways. The continuous control
loop is established to adapt to localization errors or changes. It
is noted that the configuration service may have control over the
beam-forming capabilities of the infrastructure access points. The
user device-side beam-forming adapts automatically by scanning and
selecting the best beam direction automatically using the device's
standard process.

\section{Evaluation\label{sec:Evaluation}}

\begin{figure}[tbh]
\begin{centering}
\includegraphics[width=1\columnwidth]{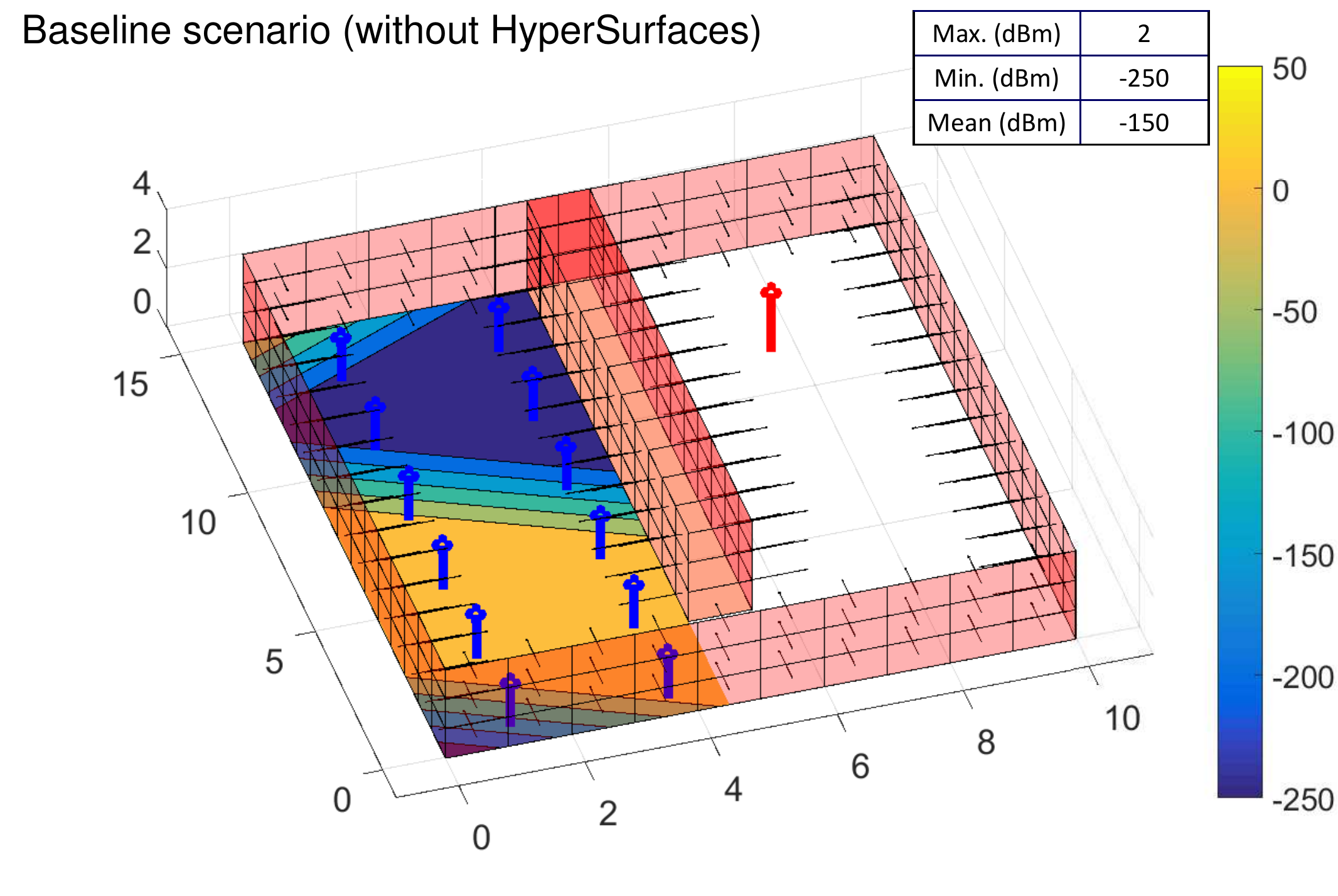}
\par\end{centering}
\begin{centering}
\includegraphics[width=1\columnwidth]{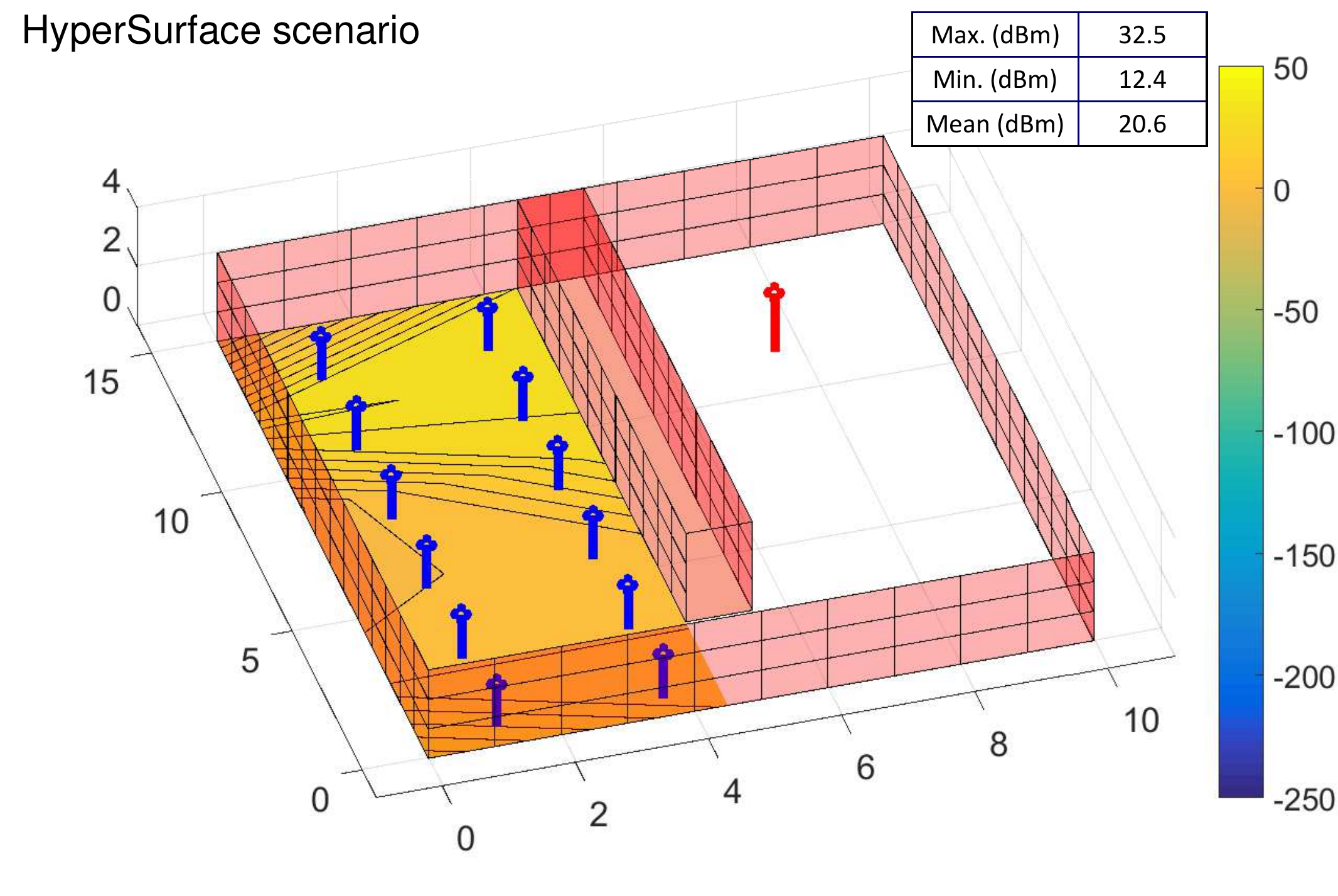}
\par\end{centering}
\begin{centering}
\includegraphics[width=1\columnwidth]{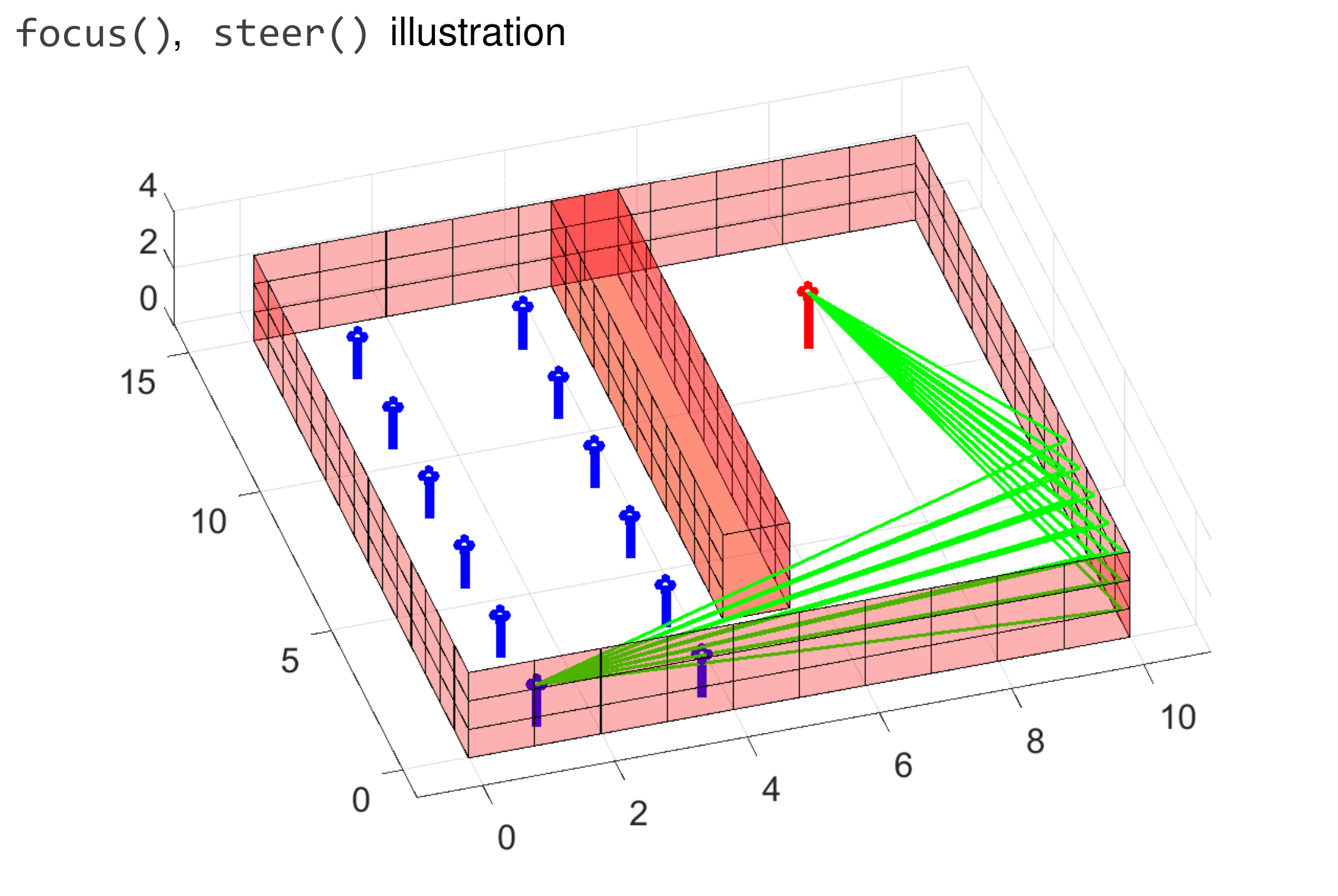}
\par\end{centering}
\caption{\label{fig:results}Simulation results for wireless environment optimization
study at $60\,GHz$. The baseline, non-HyperSurface setup (top) shows
a poor coverage. The use of HyperSurfaces (middle) shows significantly
improved signal coverage and received power overall. An illustration
of the deployed $\texttt{focus}$ and $\texttt{steer}$ functions
is shown at the bottom inset.}

\end{figure}

In this section, we present preliminary results to show the potential
of HyperSurfaces in mitigating undesired path loss effects in a real-world
wireless communication scenario. Specifically, we demonstrate the
performance improvement with the $\texttt{focus}$ and $\texttt{steer}$
function implemented in a typical indoor environment.

As shown in Fig.~\ref{fig:results}, the indoor space shows a dimension
of $15~m$ in length and $10~m$ in width and a height of $3~m$.
The room is divided by a middle wall (with a length of $12~m$ and
a thickness of $1~m$) into two sections (i.e., line-of-sight and
non-line-of-sight sections, respectively), each with a width of $4.5~m$.
All walls are coated with HyperSurface tiles with a size of $1\times1~m$.
An EM transmitter, with a height of $2~m$ as shown in red color in
Fig.~\ref{fig:results}, is located on one side of the room and equipped
with a half-dipole antenna and transmits at $60\,GHz$ with $25\,MHz$
bandwidth. The transmission power is set to $100\,dBm$. In total
$12$ receivers (shown in blue color in Fig.~\ref{fig:results})
are uniformly distributed on the non-line-of-sight side of the room
with a same height of $1.5~m$ and half-dipole antennas.

The evaluation is performed on a three-dimensional dynamic ray-tracer
developed for map-based channel modeling~\cite{terarays}, specifically
customized to implemented the $\texttt{focus}$ and $\texttt{steer}$
functionalities. This map-based ray-tracer is built based on 3D channel
models in both microwave and millimeter wave frequency bands and is
validated against field measurements. The implemented functions are
designed to be applicable to any receiver position, either in line-of-sight
(right room) or non-line-of-sight (left room) sections. In the simulated
scenario, the functions are applied to maximize the minimum received
power over the $12$ receivers in the non-line-of-sight area.

The baseline scenario, as shown in Fig.~\ref{fig:results}-top, shows
a plain setup where the norms of tiles are naturally perpendicular
to the tile surfaces without any HyperSurface functionality activated.
The average received power over the $12$ receivers is $-75\,dBm$,
while the minimum power is $-250\,dBm$ and is below the threshold
allowed by the ray-tracer, implying disconnected areas. The receivers
on the upper-right and bottom-left corners are not covered in this
setup. In comparison, with the HyperSurfaces enabled, as shown in
Fig.~\ref{fig:results}-middle, all receivers are in good coverage
with an obvious leverage of received power with an average received
power of $20.6\,dBm$. Moreover, a maximum and a minimum received
power of $32.5\,dBm$ and $12.4\,dBm$ respectively.

The work principle of tuning the HyperSurface tiles in the example
is the following: we begin with the most distant receiver (top-right
position) and assign $\texttt{focus}$ and $\texttt{steer}$ commands
to the tiles that offer the shortest air-route. The used tiles are
marked and are not used again for other receivers. We note that this
is a simplification, as metasurfaces can achieve beam splitting functionalities~\cite{Yang.2016}.
Thus, in reality, a single tile could be tuned to affect more than
one user. The process is repeated for the rest of the users. In the
context of studied, static scenario, the number of tiles to be used
for each user is deduced by a generic optimizer~\cite{optquest},
which seeks to maximize the minimum received power over all receivers.
As discussed in Section~\ref{sec:future}, however, real-time operation
is expected to require specialized optimization processes. Figure~\ref{fig:results}-bottom
provides an example of a single $\texttt{focus}$ and $\texttt{steer}$
function deployment. The tiles with green-colored paths impinged upon
will adjust their azimuth and elevation angles to focus the signals
from transmitter to desired receiver.

\section{Challenges and Research Directions\label{sec:future}}

Further research in programmable wireless environments can target
the tile architecture and the inter-tile networking, the tile control
software, and many applications such as mm-waves, D2D and 5G systems.

Regarding the tile architecture, the optimization of the dynamic meta-atom
design constitutes a notable goal towards maximizing the supported
function range of a tile. Ultra-wideband meta-atom designs able to
interact concurrently with a wide variety of frequencies, e.g., from
$1$ to $60$ GHz, constitute a notable research goal~\cite{Su.2017b}.
Formal tile sounding procedures need to be defined, i.e., simulation-based
and experimental processes for measuring the supported functions and
parameters per tile design. Additionally, the tile reconfiguration
speed needs to be studied, in order to yield the adaptivity bounds
of the programmable environments. In this sense, inter-tile networking
protocols need to be designed to offer fast, energy-efficient wireless
environment reconfiguration, supporting a wide range of user mobility
patterns. Adaptation to user mobility can also target the mitigation
of Doppler shift effects.

The HyperSurface control software needs to be optimized regarding
its complexity, modularity and interfacing capabilities. Low-complexity,
fast configuration optimizers can increase the environments maximum
adaptation speed. Towards this end, both analysis-based and heuristic
optimization processes need to be studied. Additionally, following
the Network Function Virtualization paradigm~\cite{Akyildiz.2016},
the various described and evaluated optimization objectives can be
expressed in a modular form, allowing their reuse and combination.
For example, the tiles may be configured to maximize the minimum received
power within a room, subject to delay spread restrictions. Well-defined
tile software interfaces can allow for a close collaboration with
user devices and external systems. For instance, the power delay profile
towards a user can be matched to the Multiple-Input Multiple Output
arrangement of his device. It is noted that such joint optimization
can be aligned to the envisioned 5G objectives of ultra-low latency,
high bandwidth, and support for massive numbers of devices within
an environment~\cite{Akyildiz.2016}.

We note that the HyperSurface concept is applicable to any frequency
spectrum and wireless architecture. Therefore, solving the corresponding
path loss, fading, interference and non-line-of-sight problems in
general using HyperSurfaces constitute promising research paths. Such
directions can further focus on indoors and outdoors communication
environments. In indoors settings, the HyperSurface tiles can cover
large parts of the wireless environment, such as walls, ceilings,
furniture and other objects and offer more precise control over electromagnetic
waves. In outdoors settings, the HyperSurface tiles can be placed
on key-points, such as building facades, highway polls, advertising
panels, can be utilized to boost the communication efficiency. In
both settings, i.e., in indoors and outdoors, the automatic tile location
and orientation discovery can promote the ease of deployment towards
``plug-and-play'' levels. Moreover, the joint optimization of antenna
beam-forming and tile configurations need to be studied, to achieve
the maximum performance.

Studying the use of HyperSurfaces in mm-wave systems, 5G systems and
$THz$ communications is of particular interest. For example, mm-wave
and THz systems are severely limited in terms of very short distances
and LOS scenarios. The HyperSurfaces can mitigate the acute path loss
by enforcing the lens effect and any custom reflection angle per tile,
avoiding the ambient dispersal of energy and non-line-of-sight effects,
extending the effective communication range. Dynamic meta-atoms that
can interact with $THz$ modulated waves need to be designed. This
has been shown to be possible for graphene-based metasurfaces~\cite{Lee.2012}.
The tile sensing accuracy and re-configuration speed must also match
the extremely high spatial sensitivity of $THz$ communications, calling
for novel, highly distributed tile control processes. Optical tile
inter-networking is another approach to ensure that the tile adaptation
service is fast enough for the $THz$ communication needs.

Finally, the control of EM waves via HyperSurfaces can find applications
beyond classic communications. EM interference constitutes a common
problem in highly sensitive hardware, such as medical imaging and
radar technology. In these cases, the internals of, e.g., a medical
device can be treated as an EM environment, with the objective of
canceling the interference to the imaging component caused by unwanted
internal EM scattering. Such interference can be mitigated only up
to a degree during the design of the equipment. Common discrepancies
that occur during manufacturing can give rise to unpredictable interference,
resulting into reduced equipment performance. However, assuming HyperSurface-coated
device internals, interference can be mitigated, or even negated,
after the device manufacturing, via simple software commands.

\section{Conclusion\label{sec:Conclusion}}

The present study introduced software control over the electromagnetic
behavior of a wireless environment. The methodology consisted of coating
size-able objects, such as walls, with HyperSurface tiles, a novel
class of planar materials which can interact with impinging waves
in a programmable manner. Interaction examples include wave absorbing
and steering towards custom directions. The tiles are networked and
controlled by an external service, which defines and deploys a configuration
that benefits the end-users. Notable applications are the mitigation
of propagation loss and multi-path fading effects in virtually any
wireless communication system, including mm-wave and $THz$ setups.
The study defined the HyperSurface tile architecture and the structure
of the programmable wireless environments that incorporate them. Evaluation
via simulations demonstrated the exceptional potential of this novel
concept.

\section*{Acknowledgment}

This work was partially funded by the European Union via the Horizon
2020: Future Emerging Topics call (FETOPEN), grant EU736876, project
VISORSURF (http://www.visorsurf.eu).

\bibliographystyle{IEEEtran}

\end{document}